# Mach-Zehnder interference of fractionalized electron-spin excitations

Takase Shimizu[1,2,*], Eiki Iyoda[3], Satoshi Sasaki[1], Akira Endo[2], Shingo Katsumoto[2], Norio Kumada[1], and Masayuki Hashisaka[1,2,†]

[1]*NTT Basic Research Laboratories, NTT Corporation, 3-1 Morinosato-Wakamiya, Atsugi, Kanagawa 243-0198, Japan*

[2]*Institute for Solid State Physics, The University of Tokyo, 5-1-5 Kashiwanoha, Kashiwa, Chiba 277-8581, Japan*

[3]*Department of Physics, Tokai University, Hiratsuka-shi, Kanagawa 259-1292, Japan*

email: *takase.shimizu@ntt.com, †hashisaka@issp.u-tokyo.ac.jp

Inter-channel Coulomb interaction mixes charge excitations in copropagating quantum Hall edge channels, generating coupled excitation eigenmodes propagating at different speeds. This mode transformation causes an electron state to split into fragments, corresponding to the Tomonaga-Luttinger liquid model of a chiral one-dimensional electronic system. This paper reports the coherent evolution of an electron state under the fractionalization process in a Mach-Zehnder interferometer employing copropagating spin-up and spin-down channels as the interference paths. We observe the interference visibility oscillations as a function of the voltage bias applied between the interference paths, which are attributed to the second-order interference between the fractionalized spin excitations with different phase evolutions. This observation contrasts with the single-particle picture that predicts only the first-order interference, reflecting the phase evolution of a spin-up and spin-down superposition state during the one-way transport. The second-order interference manifests the coherent splitting of the superposition state to the mutually independent fast and slow excitations. Our observation offers the fractionalization process as a novel way to encode an electron spin state to spatially separated fragments.

Quantum Hall (QH) edge states serve as chiral one-dimensional (1D) electron channels to construct electronic quantum-optical circuits [1-3]. While the circuit shows some analogies with a photonic counterpart, fermionic statistics and electron correlation often cause intriguing electron dynamics and allow us to examine the impact of the electronic nature on quantum coherence. Indeed, electron correlation drives edge states out of the single-particle picture, inducing exotic 1D electron dynamics predicted by the Tomonaga-Luttinger (TL) liquid model [4-8]. Interacting edge channels have demonstrated several exotic TL liquid behaviors, such as charge fractionalization [9] and spin-charge separation [10-13], raising the question of how an electronic state evolves coherently in such strongly correlated systems.

Coulomb interaction between copropagating spin-up and spin-down edge channels in the integer QH state mixes charge excitations in these channels, generating mutually independent fast and slow excitation eigenmodes [14]. Due to this mode transformation, a single-electron excitation is fractionalized into these eigenmodes; we refer to this process as 'electron splitting.' Previous studies identified the splitting process by injecting a spin-up or a spin-down excitation and measuring the resultant multiple excitations [10-14]. In contrast, we prepare a spin-up and spin-down superposition state to examine the coherent single-electron spin precession under fractionalization. We measured a Mach-Zehnder interferometer (MZI) employing the copropagating channels as the interference paths [15-17], which provides the best opportunity to scrutinize the coherent evolution of the superposition state [Fig. 1(a)]. In the MZI, an initial superposition state is prepared at the inter-channel tunnel coupling, i.e., beam splitter (BS), at the entrance. The superposition state precesses during the propagation toward the other BS at the exit. Then, the exit BS sends an electron to either the spin-up or the spin-down output channel, depending on the electron's spin orientation. A single-particle picture (without fractionalization) only predicts the first-order interference in the one-way interferometer. Conversely, our MZI demonstrates the feature of second-order interference, the beating of the interference visibility as a function of the voltage bias applied between the channels. The calculation using the bosonization technique [18] reveals that the second-order interference reflects the difference in the phase evolutions between the fast and slow excitations. This observation manifests coherent splitting of an electron spin state, inspiring an analogy with the Cooper-pair splitting at a superconducting junction [19].

Our MZI was fabricated in a two-dimensional electron system (2DES) at an AlGaAs/GaAs heterointerface 65 nm below the surface (electron density $n_e = 3.7 \times 10^{11}$ cm$^{-2}$ and mobility $\mu = 9 \times 10^5$ cm$^2$V$^{-1}$s$^{-1}$) [Fig. 2(a)]. We measured the sample at electron temperature $T_e = 84$ mK in a dilution refrigerator (base temperature: 43 mK). The Landau level filling factor of the bulk 2DES was set at $v \cong 4$ by a perpendicular magnetic field $B \cong 3.9$ T applied from the front to the back, leading to anticlockwise chirality of the edge channels. Large negative voltages applied to the center gate metals, $G_L$, $G_M$, and $G_R$, deplete the 2DES underneath to form the main structure of the MZI. The copropagating spin-up ($j = 1$) and spin-down ($j = 2$) channels of the lowest Landau level work as the interference paths. The sharp corners defined by $G_L$ and $G_R$ induce spin-flip scatterings between the paths through the local spin-orbit interaction, playing the role of the beam splitters BS1 and BS2, respectively [see Figs. 1(a) and 2(a)]. The inner channels [$j = 3$ and 4, see Fig. 2(b)] of the second-lowest Landau level, omitted in Fig. 2(a), are spatially far separated from the interference paths due to the large cyclotron gap and contribute minimally to the results [for details, see Supplemental Material (SM) [40]].

We applied a voltage $V_{in}$ to $j = 1$ channel (through $\Omega_1$) and measured the output current from $j = 2$ channel (through $\Omega_3$). A negative voltage to the gate metal $G_{SL}$, located immediately to the left of $G_L$, reduces the electron density underneath to form the $\nu = 1$ region. This region separates $j = 1$ and 2 channels before entering the MZI, making it possible to apply different biases to them from ohmic contacts $\Omega_1$ and $\Omega_4$, respectively. Similarly, energizing the gate metal $G_{SR}$ forms another $\nu = 1$ region that separates the output spin-up ($I_\uparrow$) and spin-down ($I_\downarrow$) currents, respectively, into $\Omega_2$ and $\Omega_3$. We measured the $V_{in}$ dependence of the differential conductance $g_\downarrow = dI_\downarrow/dV_{in}$ using the standard lock-in technique with a tiny ac modulation of $V_{in}$ (15 $\mu V_{RMS}$ at 170 Hz) (for experimental details, see Refs. [15-17].)

Figure 2(c) shows the measured $g_\downarrow$ at zero bias as a function of $B$ and $V_M$, demonstrating the Aharonov-Bohm (AB) interference. Note that $V_M$ shifts the $j = 1$ channel largely in space and the $j = 2$ channel slightly, thereby changing the area $A = wL$, where $w$ is the incompressible strip width between the channels [20], and $L$ is the path length. In the following, we discuss the AB oscillations observed by varying $V_M$ at $B = 3.9$ T. Figure 2(d) displays the measured $g_\downarrow$ plotted on the $V_M$-$V_{in}$ plane, and Fig. 2(e) shows $g_\downarrow$ traces as a function of $V_M$ at several $V_{in}$ values [indicated by arrows of corresponding colors in Fig. 2(d)]. The AB oscillations are most pronounced near $V_{in} \cong 0$ V and show nonmonotonic suppression at finite $V_{in}$, exhibiting asymmetric behaviors between the bias directions; in particular, the AB oscillation amplitude is strongly suppressed above $V_{in} \cong 200$ $\mu$V, in marked contrast to $V_{in} < -200$ $\mu$V. To quantitatively analyze the amplitude, we define the visibility $\mathcal{V}$ as $\mathcal{V} \equiv (g_\downarrow^{max} - g_\downarrow^{min})/(g_\downarrow^{max} + g_\downarrow^{min})$ at each $V_{in}$. Here, $g_\downarrow^{max}$ and $g_\downarrow^{min}$ are the maximal and minimal $g_\downarrow$ values obtained from a sinusoidal fit in $-0.45$ V $< V_M < -0.43$ V [see Fig. 2(e)]. The red line in Fig. 3(a) shows the $V_{in}$ dependence of $\mathcal{V}$, and that in Fig. 3(b) is the phase shift $\Delta\phi/\pi$ with respect to the value at $V_{in} = 0$ V. We plotted the data above $V_{in} = -140$ $\mu$V, where the systematic error in the fitting is small. Clear dips of $\mathcal{V}$ are observed at $V_{in} \cong -75$ $\mu$V and $+90$ $\mu$V, accompanied by inflections of the phase shift [indicated by arrows in Fig. 3(b)].

The observed oscillatory behavior of $\mathcal{V}$ is incompatible with the single-particle picture, which predicts no oscillations for an MZI with the same interference paths' length [21]. Moreover, it resembles that reported for the conventional MZIs, often referred to as the 'lobe structure' [22-32]. Previous studies considered two mechanisms for the visibility oscillations. The intra-channel interaction model explains the results in a graphene MZI [33,34], while the inter-channel interaction model well describes the conventional MZIs in GaAs 2DESs, where each interference path interacts with its copropagating environmental channel [35,36]. In the latter case, the inter-channel interaction induces decoherence through mode transformation, leading to the lobe structure. Our MZI is also composed of copropagating channels in a GaAs 2DES, and the inter-channel interaction dominates over the other interactions to determine the electron dynamics. The main difference, however, is that the inter-channel interaction is between the interference paths in the present setup, while it is between the main channel and the accompanying environmental channel in the conventional one. In what follows, we show the simulated bias dependence of the visibility and the phase shift in our setup and compare them with the experimental results.

The intra- and inter-channel interactions can be represented using an interaction matrix [14]

$$U = \begin{pmatrix} u_1 & u_X \\ u_X & u_2 \end{pmatrix}, \quad (2)$$

where $u_j$ ($j$ = 1 or 2) is the velocity parameter reflecting the interaction in channel $j$, and $u_X$ is the inter-channel interaction. The eigenmodes are obtained by diagonalizing $U$ as

$$S^\dagger U S = \begin{pmatrix} u_F & 0 \\ 0 & u_S \end{pmatrix}, \quad (3)$$

where

$$S = (\boldsymbol{\rho}_F, \boldsymbol{\rho}_S), \; \boldsymbol{\rho}_F = \begin{pmatrix} \rho_{F\uparrow} \\ \rho_{F\downarrow} \end{pmatrix} = \begin{pmatrix} \cos\theta \\ \sin\theta \end{pmatrix}, \text{ and } \boldsymbol{\rho}_S = \begin{pmatrix} \rho_{S\uparrow} \\ \rho_{S\downarrow} \end{pmatrix} = \begin{pmatrix} \sin\theta \\ -\cos\theta \end{pmatrix}. \quad (4)$$

Here, $u_F$ and $u_S$ are the fast- and slow-mode velocities, $\boldsymbol{\rho}_F$ is the fast eigenmode of coupled spin-up ($\rho_{F\uparrow}$) and spin-down ($\rho_{F\downarrow}$) charges, $\boldsymbol{\rho}_S$ is that of the slow mode with $\rho_{S\uparrow}$ and $\rho_{S\downarrow}$, and $\theta$ ($0 \leq \theta \leq \pi/2$) is the mixing angle. We define a parameter $\delta \equiv (\cos\theta - \sin\theta)^2 = 1 - 2\cos\theta\sin\theta$ ($0 \leq \delta \leq 1$), which reflects the difference between $u_1$ and $u_2$, namely the asymmetry between the copropagating channels. Note that $\delta$ captures the mismatch between the spin-up and spin-down charges of the fast mode, namely, the spin information carried by the fast mode. At $\delta = 0$ ($\theta = \pi/4$), the charge and spin are completely separated into the fast and slow modes, respectively, and hence, the fast mode is irrelevant to the spin precession. When $\delta$ departs from 0, the fast mode starts to carry a fragment of the spin information, which monotonically increases with $\delta$.

We simulated the AB interference in our MZI using the bosonization technique. The mathematical formula for the calculation is described in the joint paper of this letter [18]. Figure 4(a) shows the dependence of the calculated visibility $\mathcal{V}_{\text{calc}}$ on the dimensionless chemical-potential difference $\overline{\Delta\mu} \equiv \Delta\mu/h \times L/u_S$ for several $\delta$ values. The visibility oscillations are observed at finite $\delta$, which declines with decreasing $\delta$ and ends up with the non-oscillating $v_{\text{calc}} = 1$ line at $\delta = 0$. The $\mathcal{V}_{\text{calc}}$ oscillations reflect the second-order interference of the eigenmodes $\boldsymbol{\rho}_F$ and $\boldsymbol{\rho}_S$ with the phase difference $\sim 2\pi\overline{\Delta\mu}$ between them caused by the difference in their speeds ($u_F \gg u_S$ in the integer QH state [13]).

The suppression of the oscillation amplitude with decreasing δ reflects the reduction of the spin information carried by the fast mode. Because the fast mode at δ = 0 does not have information about which interference path an electron exists on (the 'which-path' information), it does not contribute to the interference. Figure 4(b) displays the phase shift $\Delta\phi_{calc}/\pi$ with respect to the value at $V_{in}$ = 0 V. When δ is finite, $\Delta\phi_{calc}/\pi$ monotonically increases with $\overline{\Delta\mu}$ near zero bias and oscillates around constant values in the high positive and negative bias regions beyond the $\mathcal{V}_{calc}$ dip positions. The linear phase shift near zero bias reflects the velocity difference ($u_1$ and $u_2$) between the eigenmodes [18]. Note that in the above calculations, we assumed that BSs have the same transmission probabilities ($T_1 = T_2 \equiv T$): in this case, the visibility of the $g_\downarrow$ oscillations is independent of $T$, in contrast to the case of measuring $g_\uparrow = dI_\uparrow/dV_{in}$, where the change in $T$ varies the visibility [22].

The $\mathcal{V}_{calc}$ and $\Delta\phi_{calc}/\pi$ traces at finite δ [Fig. 4] capture major features in the experimental results [Fig. 3]. Both simulation and experimental data show clear dips in visibility at finite bias. They also show similar monotonical increases in the phase shifts near zero bias and their inflections near the visibility dip positions. The dips are interpreted as indicative of the destructive interference between the fast and slow excitations. The gradual decrease of $\mathcal{V}_{calc}$ accompanied by the oscillations with increasing bias is signature of the long-range correlation reflecting the TL liquid nature of the copropagating channels, similar to the conventional MZI [33]. Thus, the simulation at finite δ, taking only the dominant inter-channel interaction into account, successfully represents several experimental features and provides their interpretations. Indeed, the experimental device [Fig. 2(a)] would have δ > 0 since Coulomb couplings of $j$ = 1 and 2 channels with the surroundings naturally differ from each other [13]. From these findings, we conclude that the visibility oscillations observed in the experiment result from the second-order interference of fractionalized spin excitations, manifesting the coherent splitting of the superposition spin state.

The comparison of the visibility dip positions between the simulation and experiment allows us to conduct a quantitative analysis. In our theoretical model [18], the fast and slow excitations gain the phase difference $2\pi\Delta\mu/h \times L/u_S$ at the MZI exit. Therefore, we can extract the slow-mode velocity $u_S$ ~ 3 × 10$^4$ m/s from either of the measured dip values $V_{in} \cong$ −75 μV or +90 μV. The $u_S$ value roughly agrees with the slow-mode velocities reported in previous studies for similar samples [13, 37] and the velocity (~1.5 × 10$^4$ m/s) estimated from the numerical evaluation of matrix $U$ elements using the distributed circuit model [13, 38, 39]. The difference in the dip positions between the positive and negative bias directions suggests that the edge electron-density profile and hence, the slow-mode velocity, depend on the bias. Note that while the interaction between the interference paths and the environmental $j$ = 3 and 4 channels can modify the lobe pattern of the visibility, the contribution is minor because these channels are spatially far separated [40].

The coherence of the electron splitting process suggests quantum entanglement between the fractionalized excitations. We evaluated von Neumann entanglement entropy $S_N$ between the fast and slow modes transformed from the superposition state generated at $BS_1$ [18]. When the copropagating channels are asymmetric ($\delta > 0$), $S_N$ is positive at $T \neq 0$ and $T \neq 1$ and increases to $S_N = \log 2$ with $T$ approaching 1/2. This observation indicates that a tunneling process at the BS and the subsequent coherent mode transformation result in the entanglement between the fractionalized excitations, underscoring the fundamental capability of interacting 1D channels to generate spatially separated entangled pairs (for details, see [18].)

While several experimental features can be explained by considering only the inter-channel interaction between the interference paths as discussed above, further detailed comparison between theory and experiment shows qualitative differences. The main reason for the lack of close agreement may be the perturbative treatment for the BS transmission probability in the simulations [18]. In addition, our simulation neglects many factors present in the actual system; for example, a possible difference between BS1 and BS2, a small but finite interaction of the interference paths with the environmental $j = 3$ and 4 channels, and the effect of thermal noise, which smears the visibility oscillations at high biases [26,29,36]. We leave a more quantitative analysis, including these issues, for future studies.

Lastly, we discuss the significant source of decoherence in the present system. Figures 5(a)-(c) show schematic energy diagrams of the $v = 1$ and 2 Landau levels near the 2DES edge under different bias conditions. Whereas the spin-up and spin-down channels are distant at negative [Fig. 5(a)] and low positive bias [Fig. 5(b)], they spatially overlap [Fig. 5(c)] at high positive bias to enhance the inter-channel inelastic scattering [41]. The inelastic process occurs everywhere along the copropagating channels as well as at BSs [Fig. 5(d)], causing strong dissipation to suppress the interference visibility. We examined this scenario by measuring the current noise in a control device on the same sample chip, a BS (BSc) having the same corner angle as BS1 and BS2 [see the inset of Fig. 5(f)]. For details of this experimental setup, see SM [40]. Figures 5(e) and (f) show the simultaneously measured bias dependence of the differential conductance $g_C \equiv dI_\downarrow^C/dV_{in}^C$ and the current noise cross-correlation $S_X = \langle \Delta I_\uparrow^C \Delta I_\downarrow^C \rangle$, respectively [42-45]. We observe that $g_C$ varies with the input voltage $V_{in}^C$, and that $S_X$ decreases from zero to show negative cross-correlation between the two outputs. At $V_{in}^C < 110$ μV, $S_X$ agrees well with the theoretical curve (black curve) for the shot noise generation at BSc [45],

$$S_{calc} = -2e|I_\downarrow^C|F\left[\coth\left(\frac{eV_{in}^C}{2k_BT_e}\right) - \frac{2k_BT_e}{eV_{in}^C}\right], \quad (5)$$

where $I_\downarrow^C$ is the output spin-down current, $F = 1 - T_C$ is the noise reduction factor, and $k_B$ is Boltzmann's constant. Here, we evaluated the transmission probability of BSc as $T_C = I_\downarrow^C/V_{in}^C \times h/e^2$. Meanwhile, the shot noise is strongly suppressed ($|S_X| < |S_{calc}|$) at $V_{in}^C > 110$ μV, indicating the inelastic scattering illustrated in Fig. 5(c) [46]. The threshold voltage of the noise reduction, $V_{in}^C \cong 110$ μV, corresponds to the bias where $g_C$ rapidly increases to approach $e^2/2h$ [Fig. 5(e)], which implies charge equilibration between the copropagating channels. It also coincides with the onset bias $V_\text{in}$, above which the interference visibility is strongly suppressed [Figs. 2(d) and 3(a)]. These synchronized behaviors unambiguously reveal the impact of the inelastic scattering as the decoherence mechanism.

We have identified the coherent spin fractionalization process in the bias dependence of the visibility and the phase shift of the AB oscillations. We have also demonstrated that the inelastic inter-channel scattering is the source of decoherence. In contrast to the conventional MZIs, where inter-channel interaction induces the oscillatory visibility behavior through the decoherence process, our experiment illuminates the complementary aspect, the coherent evolution of a single-electron spin state under fractionalization. Our results highlight Coulomb interaction as an essential tuning knob for advanced spin manipulations in electronic quantum-optical circuits. As a prospect, spin fractionalization in coupled MZIs [47] and a collision experiment [48] between entangled fragments may provide novel functionalities for manipulating spin excitations in condensed matter.

The authors thank K. Muraki and T. Fujisawa for fruitful discussions and H. Murofushi and M. Imai for technical support. This work was supported by Grants-in-Aid for Scientific Research (Grant Nos. JP19H05603, JP22H00112, and JP24H00827) and JSPS Bilateral Program Number JPJSBP120249911.

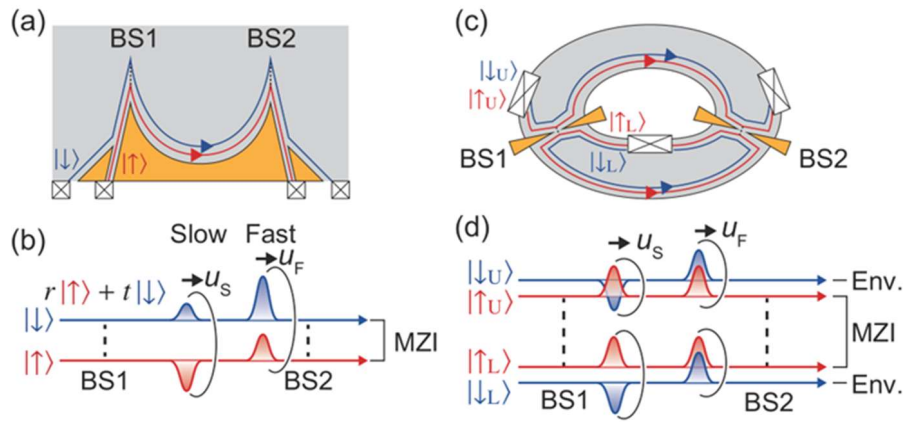

Fig. 1. (a)(b) MZI employing copropagating spin-up (red) and spin-down (blue) channels as the interference paths. Sharp corners operate as BS1 and BS2. BS1 prepares the initial superposition state, which precesses during the propagation toward BS2. The superposition state splits into the fast and slow modes with the velocities $u_F$ and $u_S$, respectively. (c)(d) Conventional (spinless) MZI. Quantum point contacts (QPCs) work as BS1 and BS2. The spin-down channels copropagating with the paths behave as the environment through the inter-channel interaction. The interaction between the interference paths is negligible in this case.

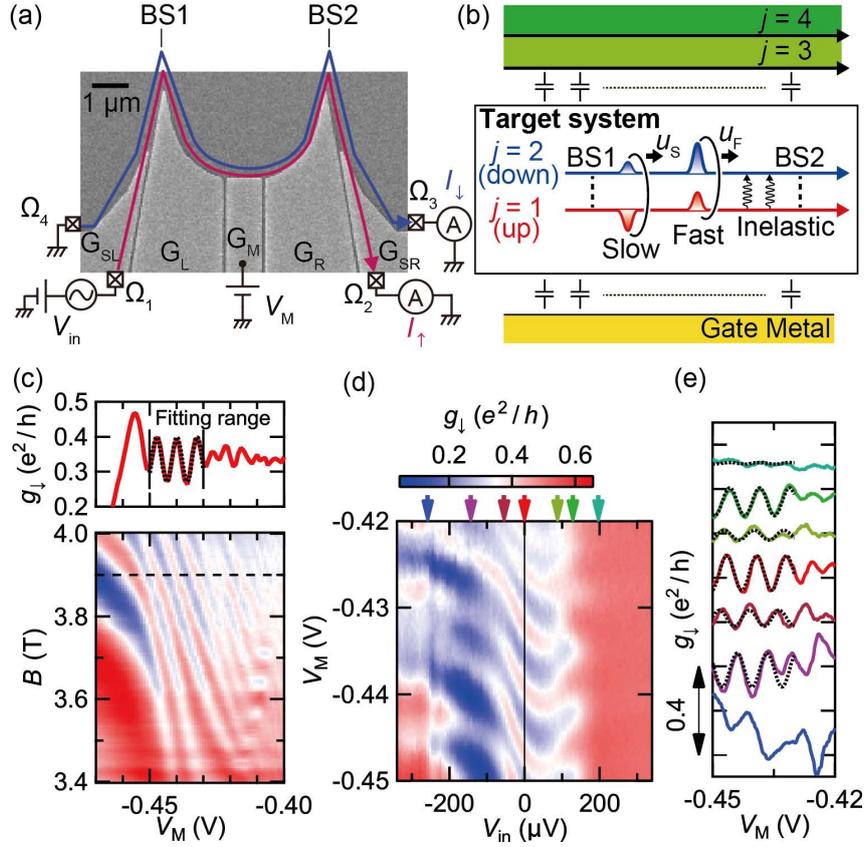

Fig. 2. (a) Scanning electron micrograph of the MZI and measurement setup. (b) Target system constructed using spin-up ($j = 1$) and spin-down ($j = 2$) channels. Inner $j = 3$ and 4 channels and the gate metals behave as environment through weak Coulomb interaction, inducing the asymmetry between interference paths. (c) Bottom: color plot of measured $g_\downarrow$ as a function of $B$ and $V_M$. Top: Cross section of the color plot at $B = 3.9$ T, indicated by dashed line in the bottom panel. (b) Measured $g_\downarrow$ as a function of $V_M$ and $V_{in}$ at $B = 3.9$ T. (e) $g_\downarrow$ vs. $V_M$ at several $V_{in}$ values indicated by arrows of corresponding colors in (d). Dashed lines show sinusoidal fit.

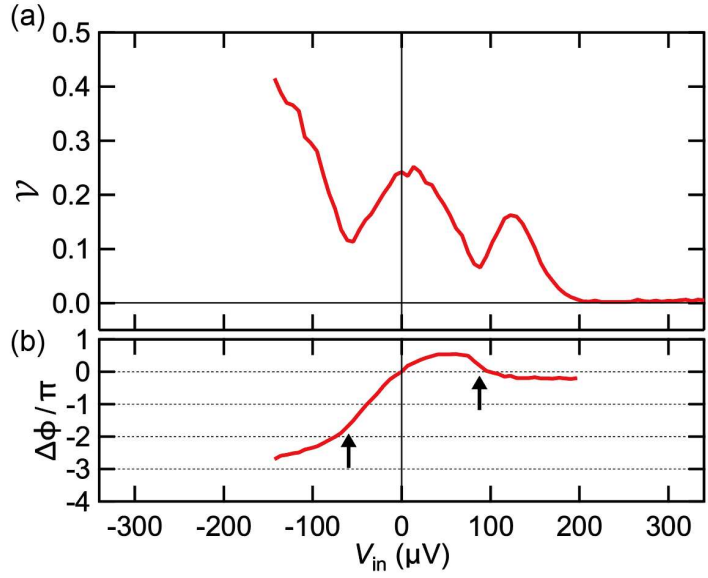

Fig. 3 Bias dependence of (a) $\mathcal{V}$ and (b) $\Delta\phi/\pi$ obtained by the sinusoidal fitting of the AB interference patterns in Fig. 2 (e) over the range of $-0.45$ V $< V_M < -0.43$ V.

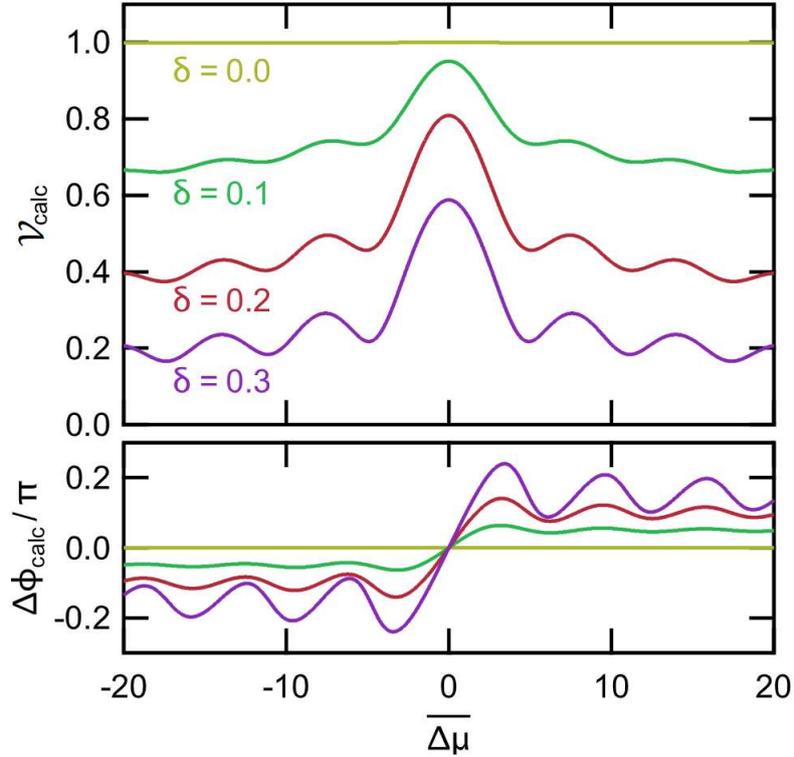

Fig. 4 Calculated visibility $\mathcal{V}_{\text{calc}}$ and phase shift $\Delta\phi_{\text{calc}}/\pi$ as a function of dimensionless chemical potential difference $\overline{\Delta\mu}$ at several $\delta$ values.

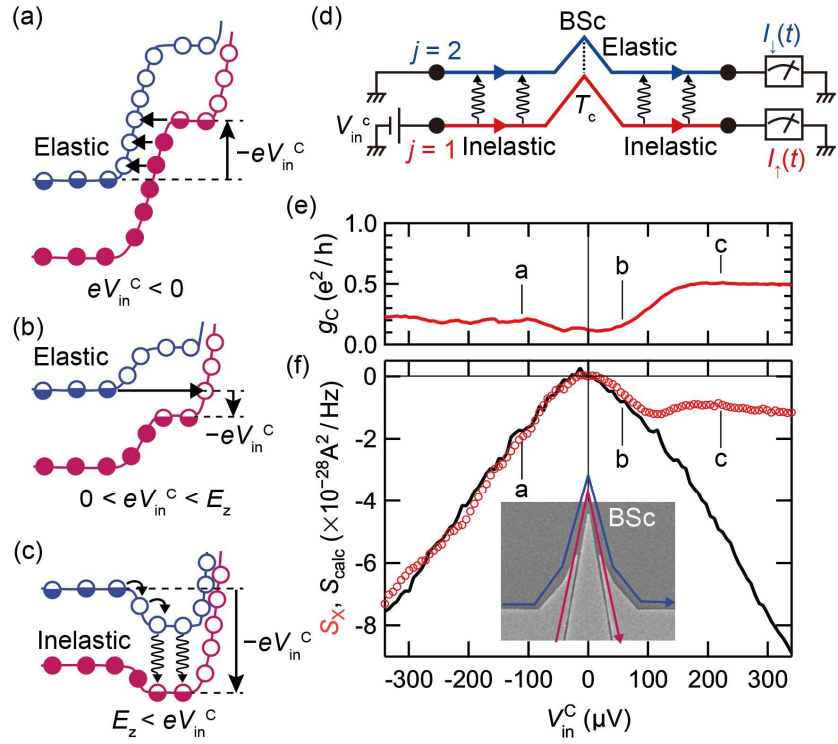

Fig. 5 Inelastic inter-channel scatterings. (a-c) Schematic of edge states at several $V_{in}$. (d) Schematic of the control experiment. (e) Bias dependence of the differential conductance. (f) Measured current noise cross correlation $S_X$ (red circles) as a function of $V_{in}^C$. Solid curve is the simulation using Eq. (5). (Inset) Scanning electron micrograph of $BS_c$.

# Supplemental material for
# "Mach-Zehnder interference of fractionalized electron-spin excitations"

Takase Shimizu[1,2,*], Eiki Iyoda[3], Satoshi Sasaki[1], Akira Endo[2], Shingo Katsumoto[2], Norio Kumada[1], and Masayuki Hashisaka[1,2,†]

[1]*NTT Basic Research Laboratories, NTT Corporation, 3-1 Morinosato-Wakamiya, Atsugi, Kanagawa 243-0198, Japan*

[2]*Institute for Solid State Physics, The University of Tokyo, 5-1-5 Kashiwanoha, Kashiwa, Chiba 277-8581, Japan*

[3]*Department of Physics, Tokai University, Hiratsuka-shi, Kanagawa 259-1292, Japan*

email: *takase.shimizu@ntt.com, †hashisaka@issp.u-tokyo.ac.jp

## 1. Measurement setup for the control experiment.

Fig. S1 shows the measurement setup for the control experiment on a single beam splitter (BS: BSc). The electron density underneath the metal electrodes $G_{SLc}$ and $G_{SRc}$ are decreased to form the $v = 1$ QH state to separate the spin-up and the spin-down channels spatially. Copropagating channels are formed in the middle region along the metal electrode $G_{BSc}$. Elastic spin-flip tunneling occurs at the sharp corner of $G_{BSc}$. To perform the noise measurement, we applied a source-drain voltage $V_{in}^c$ to contact $\Omega_1$. We measured the current noise $\Delta I_2$ at contact $\Omega_2$ and $\Delta I_4$ at contact $\Omega_4$ simultaneously with the dc outputs $I_3$ and $I_5$ and evaluated the cross-correlation $S_X = \langle \Delta I_\uparrow \Delta I_\downarrow \rangle = \langle \Delta I_2 \Delta I_4 \rangle$.

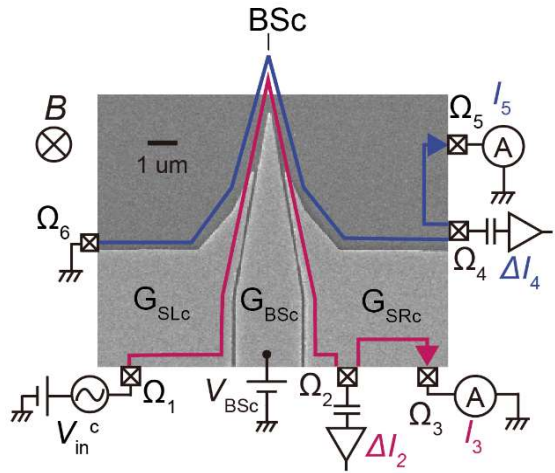

Fig. S1. Scanning electron micrograph of BSc and measurement setup. Red and blue arrows indicate the spin-up and spin-down channels in the lowest Landau level ($n_0$ channels.) The inner channels ($n_1$ channels) are omitted in this figure.

## 2. Inner $j = 3$ and $4$ channels

The inner $j = 3$ and $4$ channels (here, we refer to them $n_1$ channels) are spatially far separated from the main $j = 1$ and 2 channels ($n_0$ channels) of the MZI due to the large cyclotron gap. In this case, the Coulomb interaction between $n_0$ and $n_1$ channels is smaller than the interaction between $j = 1$ and 2 channels, and, therefore, the inner channels give a minor contribution to the AB oscillations. The far separation can be confirmed by measuring the tunneling current between $n_0$ and $n_1$ channels in the control device (Fig. S1), for example.

Figure S2 shows the bias dependence of the differential conductance $g_c \equiv dI_5^C/dV_{in}^C = t_C \times e^2/h$ measured at the filling factor of the ungated region at $v = 4$ and the local filling factors underneath $G_{SLc}$ and $G_{SRc}$ at $v = 1$ and $v = 2$, respectively. In this case, $t_C$ means the tunneling probability between $j = 1$ and $j = 3$ channels, which reflects their distance. The data shows that $t_C$ remains very small (below 2%) over the entire range of the bias despite the same spin orientation between the channels. This fact contrasts with the case of tunneling between $j = 1$ and 2 channels with larger $t_C$ above 10% (dotted line: see main text for details), even though their spin orientations differ.

We also calculated the spatial distance between the edge channels (the width of the incompressible strips) using the classical electrostatic model for the QH edge states [1]. The model gives the width of the $i$th incompressible strip as

$$w_i \approx \sqrt{\frac{8|\Delta E_i|\epsilon\epsilon_0}{\pi e^2 (dn/dr)|_{r=r_i}}},$$

where $\Delta E_i$ is the energy gap between $j = i$ and $i + 1$ Landau levels, $\epsilon\epsilon_0$ is the dielectric permittivity, $n(r)$ is the electron density, and $r_i$ is the center of the $i$th incompressible strip from the edge of the gate metal. Assuming $\Delta E_1 \approx 0.10$ meV for the exchange-enhanced gap between $j = 1$ and 2 states and $\Delta E_2 \approx 6.7$ meV for the cyclotron gap at $B = 3.9$ T, we obtain the estimations of $w_1 = 1.4$ nm and $w_2 = 19$ nm within the frozen-surface-charge model [1].

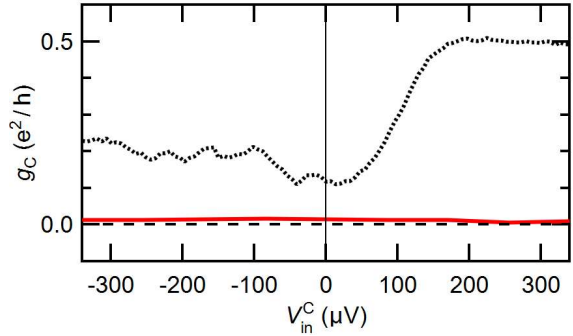

Fig. S2. Bias dependence of the differential conductance $g_c$ between $j = 1$ and 3 channels (red). Dotted line indicates $g_c$ between $j = 1$ and 2 channels [same data as that in Fig. 5(e)].

From these observations, we conclude that $n_1$ channels are far separated from the $n_0$ channels, and the electron dynamics in the present system is dominated by the inter-channel interaction between $j = 1$ and 2 channels.